\documentclass[amsmath,amssymb,amsfonts,aps,pre,preprint,superscriptaddress,bibnotes,showpacs,showkeys,longbibliography]{revtex4-1}

\usepackage{graphicx}
\usepackage{subfigure}
\usepackage{amsmath}
\usepackage{amsfonts}
\usepackage{amssymb}
\usepackage{natbib}
\usepackage{relsize}
\usepackage{sidecap}
\usepackage{braket}
\usepackage{footmisc}
\usepackage{footnote}
\usepackage{color}
\begin{document}

\title{Casimir force of two-component Bose-Einstein condensates confined by a parallel plate geometry}

\author{Nguyen Van Thu}
\affiliation{Institute for Research and Development, Duy Tan University, Da Nang, Vietnam}
\affiliation{Department of Physics, Hanoi Pedagogical University 2, Hanoi, Vietnam}

\begin{abstract}
Using field theory we calculate the Casimir energy and Casimir force of two-component Bose-Einstein condensates restricted between two parallel plates, in which Dirichlet and periodic boundary conditions applied. Our results show that, in one-loop approximation, the Casimir force equals to summation of the one of each component and it is vanishing in some cases: (i) inter-distance between two plates becomes large enough; (ii) intraspecies interaction is zero; (iii) interspecies interaction is full strong segregation.
\end{abstract}

\maketitle

\section{Introduction}\label{sec:1}

Approximately 70 years have passed since H. B. G. Casimir published his famous paper \cite{Casimir}, which mentioned a new kind of the force caused by the distorted vacuum fluctuation of quantized electromagnetic field and called Casimir force, there are many attentions to research the Casimir force in both experimental and theoretical sides in many scopes of physics: massless scalar field \cite{Edery}, quark matter \cite{PhatThu}, quantum liquids \cite{Recati}, nanotechnology and many others \cite{Bordag}.

In field of Bose-Einstein condensate (BEC), the Casimir force is considered as the quantum fluctuations on top of ground state, which corresponds to phononic excitations \cite{Biswas, Biswas2,Biswas3,Schiefele}. For the noninteracting BEC the Casimir force is vanished \cite{Biswas, Biswas2} and in limit of weakly interacting BEC this force is nonzero \cite{Biswas3,Schiefele}. Besides, the Casimir force in BEC at finite temperature was considered in Refs. \cite{Biswas, Biswas2,Martin}. Although the Casimir force of a BEC has not been measured yet but a successful for Casimir-Polder force in both theory \cite{Casimir2} and experiment \cite{Harber} are the motivation for physicists in BEC field.

To our understanding, the study of the Casimir force in two-component Bose-Einstein condensates (BECs) is so far still absent although many aspects have been researched, for example, statics properties \cite{Thu,Joseph,Deng}, dynamical excitations \cite{Mazets,Thu2}, {\it etc}. The goal of this paper is to remedy this gap. To do this we consider a BECs is confined to a parallel plate geometry with the size $L_x, L_y$ and inter-distance is $L=L_z$, which satisfies condition $L_x,L_y\gg L$. This means that our system is limited in the volume $V=L_xL_yL$ as was discussed in \cite{Lipowsky}. Here we assume that two plates perpendicular to $z$-axis.

This paper is organized as follow. In Section \ref{sec:2} we investigate the Casimir energy and Casimir force of BECs. The conclusions are  given in Section \ref{sec:3} to close the paper.

\section{Casimir force of Bose-Einstein condensate mixtures \label{sec:2}}

Let us begin with the idealized binary mixture of Bose gases, without external field, given by the Lagrangian density \cite{Phat,Pethick},
\begin{eqnarray}
{\cal L}=\sum_{j=1,2}\left[\frac{i\hbar}{2}(\psi_j^*\partial_t\psi_j-\psi_j\partial_t\psi_j^*)+\frac{\hbar^2}{2m_j}|\nabla\psi_j|^2\right]-{\cal V},\label{Lagrangian}
\end{eqnarray}
in which, the interaction potential has the form
\begin{eqnarray}
{\cal V}=\sum_{j=1,2}\left(-\mu_j|\psi_j|^2+\frac{g_{jj}}{2}|\psi_j|^4\right)+g_{12}|\psi_1|^2|\psi_2|^2,\label{potential}
\end{eqnarray}
where, for species $j$, $\psi_j=\psi_j(\vec{r},t)$ is the wave function of the condensate, which plays the role of order parameter, $m_j, \mu_j$ are atomic mass and chemical potential, respectively. The strength of repulsive inter- and intraspecies interaction determined by $g_{jj}=4\pi\hbar^2a_{jj}/m_j>0$ and $g_{12}=2\pi\hbar^2a_{12}(1/m_1+1/m_2)>0$, and $a_{jj'}$ is the $s$-wave scattering length, relevant at low energies. We consider here both cases: the system is miscible when $g_{12}^2-g_{11}g_{22}<0$ and immiscible if $g_{12}^2-g_{11}g_{22}>0$ \cite{AoChui}.

By setting $\psi_j(\vec{r},t)=\Psi_j(\vec{r}) e^{-i\mu_jt/\hbar}$ one can obtain the time-independent Gross-Pitaevskii (GP) equations
\begin{subequations}
\begin{eqnarray}
&&-\frac{\hbar^2}{2m_1}\nabla^2\Psi_1-\mu_1\Psi_1+g_{11}|\Psi_1|^2\Psi_1+g_{12}|\Psi_2|^2\Psi_1=0,\label{GPa}\\
&&-\frac{\hbar^2}{2m_2}\nabla^2\Psi_2-\mu_2\Psi_2+g_{22}|\Psi_2|^2\Psi_2+g_{12}|\Psi_1|^2\Psi_2=0.\label{GPb}
\end{eqnarray}\label{GP}
\end{subequations}

Now we invoke field theory to consider. In tree-approximation, the inverse propagators correspond two fields are
\begin{eqnarray}
D_1^{-1}(k)&=&\left(
              \begin{array}{cc}
                \frac{\hbar^2\vec{k}^2}{2m_1} & -\omega \\
                \omega &  \frac{\hbar^2\vec{k}^2}{2m_1} +g_{11}\Psi_1^2 \\
              \end{array}
            \right),\nonumber\\
D_2^{-1}(k)&=&\left(
              \begin{array}{cc}
                \frac{\hbar^2\vec{k}^2}{2m_2} & -\omega \\
                \omega &  \frac{\hbar^2\vec{k}^2}{2m_2} +g_{22}\Psi_2^2 \\
              \end{array}
            \right).\label{propagators}
\end{eqnarray}
and the density profiles correspond to minimum of potential (\ref{potential}), namely,
\begin{subequations}
\begin{eqnarray}
&&-\mu_1\Psi_1+g_{11}|\Psi_1|^2\Psi_1+g_{12}|\Psi_2|^2\Psi_1=0,\label{treea}\\
&&-\mu_2\Psi_2+g_{22}|\Psi_2|^2\Psi_2+g_{12}|\Psi_1|^2\Psi_2=0,\label{treeb}
\end{eqnarray}\label{tree}
\end{subequations}
yielding
\begin{eqnarray}
|\Psi_1|^2=\frac{g_{22}\mu_1-g_{12}\mu_2}{g_{11}g_{22}-g_{12}^2},~|\Psi_2|^2=\frac{g_{11}\mu_2-g_{12}\mu_1}{g_{11}g_{22}-g_{12}^2}.\label{profiles}
\end{eqnarray}
It is easily to see that, in tree-approximation we get the profiles (\ref{profiles}), which coincides with the one of Thomas-Fermi approximation applied for GP theory \cite{Pitaevskii}. In this approximation, Bogoliubov dispersion relation for condensate $j$ reads as
\begin{eqnarray}
E_j(k)=\sqrt{\frac{\hbar^2k^2}{2m_j}\left(\frac{\hbar^2k^2}{2m_j}+g_{jj}|\Psi_j|^2\right)}.\label{dispersion}
\end{eqnarray}
In long wavelength limit, Eq. (\ref{dispersion}) reduces to
\begin{eqnarray}
E_j(k)\approx k\sqrt{\frac{\hbar^2}{2m_j}g_{jj}|\Psi_j|^2},\label{phonon}
\end{eqnarray}
which corresponds to Goldstone bosons due to $U(1)\times U(1)$ breaking. This quasi-particles propagate in the same way as photon of electromagnetic field in original Casimir's calculations \cite{Casimir}, of cause, in our case, the velocity of quarsi-particles equals to the 'sound velocity'.

Next we consider the system in one-loop approximation. In this limit, the thermodynamical potential can be written as \cite{Andersen},
\begin{eqnarray}
\Omega={\cal V}+\sum_{j=1,2}\frac{1}{2}\int_\beta \mbox{Tr}\ln D_j^{-1}(k),\label{1loop}
\end{eqnarray}
where we used the notation
\begin{eqnarray*}
\int_\beta f(k)=T\sum_{n=-\infty}^{+\infty}\int \frac{d^3\vec{k}}{(2\pi)^3}f(\omega_n,\vec{k}),
\end{eqnarray*}
and $\omega_n$ is Matshubara frequency. For the first term in right hand side of (\ref{1loop}), combining (\ref{potential}) and (\ref{profiles}) one get
\begin{eqnarray}
{\cal V}=\frac{g_{11}\mu_2^2-2g_{12}\mu_1\mu_2+g_{22}\mu_1^2}{2(g_{12}^2-g_{11}g_{22})}.\label{F0}
\end{eqnarray}
After renormalization, the chemical potentials can be found based on (\ref{1loop}) by taking derivative the the thermodynamics with respect to density particle $n_j=|\Psi_j|^2$,
\begin{eqnarray*}
\mu_j=-\frac{\partial\Omega}{\partial n_j}.
\end{eqnarray*}
These equations lead to
\begin{eqnarray}
\mu_1&=&g_{11}n_1+g_{12}n_2,\nonumber\\
\mu_2&=&g_{22}n_2+g_{12}n_1.\label{chemical}
\end{eqnarray}
Hence, reinserting (\ref{chemical}) into (\ref{F0}) one obtains the energy density
\begin{eqnarray}
{\cal E}=\frac{1}{2}(g_{11}n_1^2-2g_{12}n_1n_2+g_{22}n_2^2).\label{energydensity}
\end{eqnarray}
It is obviously that this result coincides to  well-known result of Lee and Yang for single component Bose gas \cite{Lee}.

Making summation over $\omega_n$, the second term in right hand side (\ref{1loop}) has form
\begin{eqnarray}
\frac{1}{2}\int_\beta  {\text{Tr}}\ln D_j^{-1}(k)=\frac{1}{2}\int\frac{d^3\vec{k}}{(2\pi)^3}\big[E_j+2T\ln(1-e^{-E_j/k_BT})\big],\label{secondterm}
\end{eqnarray}
with $T$ is temperature and $k_B$ be Boltzmann constant. Note that we are studying on Casimir energy and Casimir force at zero-temperature caused by quantum fluctuation, therefore the temperature-dependence term can be dropped out of (\ref{secondterm}). The free energy in one-loop approximation (\ref{1loop}) can be rewritten as the one for single component Bose gas \cite{Andersen},
\begin{eqnarray}
\Omega={\cal V}+\Omega_1={\cal V}+\sum_{j=1,2}\Omega_{1j},\label{energy}
\end{eqnarray}
with
\begin{eqnarray}
\Omega_{1j}=\frac{1}{2}\int\frac{d^3\vec{k}}{(2\pi)^3}\sqrt{\frac{\hbar^2k^2}{2m_j}\bigg(\frac{\hbar^2k^2}{2m_j}+g_{jj}\Psi_j^2\bigg)}+\Delta_{1j}\Omega.\label{term1}
\end{eqnarray}
Here $\Delta_{1j}\Omega$ is one-loop counter term, which cancels the ultraviolet divergences \cite{Andersen}.

For sake of simplicity, we introduce healing length $\xi_j=\hbar/\sqrt{2m_j n_{j0}g_{jj}}$ as the unit of length, dimensionless wavelength $\kappa_j=k\xi_j$ and the order parameters are scaled $\phi_j=\Psi_j/\sqrt{n_{j0}}$ with $n_{j0}$ is bulk density of component $j$. In this respect, Eq. (\ref{term1}) has the form
\begin{eqnarray}
\Omega_{1j}=\frac{g_{jj}n_{j0}}{2\xi_j^3}\int\frac{d^3\vec{\kappa}_j}{(2\pi)^3}\sqrt{\kappa_j^2(\kappa_j^2+\phi_j^2)}.\label{term11}
\end{eqnarray}
Now let us investigate the Casimir energy of BECs confined between a pair of parallel plates. Let $L_j$ being inter-distance (in unit of healing length) of double plates, we have to quantize the momentum component perpendicular to the plates as follows
\begin{eqnarray}
\kappa_j^2\rightarrow \kappa_{j\perp}^2+\kappa_{jn}^2,~\kappa_{jn}=\frac{2\pi n}{L_j}\equiv\frac{n}{\overline{L}_j},~\overline{L}_j=\frac{L_j}{2\pi},~n\in{\mathbb{Z}}.\label{quantize}
\end{eqnarray}
This means that the periodic boundary condition is employed in Eq. (\ref{quantize}). Correspondingly, the integral over momentum perpendicular to the plates is replaced by a summation
\begin{eqnarray}
\int\frac{d^3\kappa_j}{(2\pi)^3}\rightarrow\sum_{n=-\infty}^\infty \int\frac{d^2\kappa_{j\perp}}{(2\pi)^2}.\label{change}
\end{eqnarray}
Applying Eqs. (\ref{change}) and (\ref{quantize}) into (\ref{term11}) we arrive
\begin{eqnarray}
\Omega_{1j}=\frac{g_{jj}n_{j0}}{2\xi_j^2\overline{L}_j^2}\sum_{n=-\infty}^\infty\int\frac{d^2\kappa_{j\perp}}{(2\pi)^2}\sqrt{(\overline{L}_j^2\kappa_{j\perp}^2+n^2)(M_j^2+n^2)},\label{term12}
\end{eqnarray}
with
\begin{eqnarray}
M_j=\overline{L}_j\sqrt{\kappa_{j\perp}^2+\phi_j^2}.\label{massj}
\end{eqnarray}
As pointed out in Refs. \cite{Saharian,Schiefele}, within Dirichlet boundary condition applied to parallel plates, Casimir energy related to the free energy in one-loop approximation for each unit area satisfies
\begin{eqnarray}
\overline{\Omega}_1=L\Omega_1\bigg|_V+\sum_j{\cal E}_{Cj},\label{2.18}
\end{eqnarray}
in which
\begin{eqnarray}
{\cal E}_C=\sum_{j=1,2}{\cal E}_{Cj},\label{definition}
\end{eqnarray}
is called Casimir energy. The summation in (\ref{term12}) can be evaluated by using Abel-Plana formula \cite{PhatThu},
\begin{eqnarray}
\sum_{n=0}^\infty f(n)=\int_0^\infty dx f(x)+\frac{1}{2}f(0)+i\int_0^\infty dx\frac{f(ix)-f(-ix)}{e^{2\pi x}-1}.\label{Abel}
\end{eqnarray}
Note that the first term in (\ref{Abel}) is divergent but it is absorbed by the counter-term. The term $f(0)$ in (\ref{Abel}) cancels out. The result is
\begin{eqnarray}
{\cal E}_{Cj}=-\frac{2g_{jj}n_{j0}}{\xi_j^2\overline{L}_j^2}\int_0^\infty  \frac{d^2\kappa_{j\perp}}{(2\pi)^2}\int_{\overline{L}_j\kappa_{j\perp}}^{M_j}\frac{\sqrt{(x^2-\overline{L}_j^2\kappa_{j\perp}^2)(M_j^2-x^2)}}{e^{2\pi x}-1}dx.\label{Casimir}
\end{eqnarray}
By changing the order of integrations in (\ref{Casimir}), the $\kappa_{j\perp}$-integral can be performed and we express (\ref{Casimir}) in the form
\begin{eqnarray}
{\cal E}_{Cj}=\int_0^\infty dx \frac{\rho_j(x,\overline{L}_j)}{e^{2\pi x}-1},\label{energyfinal}
\end{eqnarray}
with $\rho_j(x,\overline{L}_)$ is density of state function for component $j$, which has the form
\begin{eqnarray}
\rho_j(x,\overline{L}_j)=\left\{
                         \begin{array}{ll}
                           -\frac{g_{jj}n_{j0}}{8 \pi  \overline{L}_j^4 \xi_j ^2}\bigg[x \sqrt{\overline{L}_j^2 \phi_j ^2-x^2} (2 x^2-\overline{L}_j^2 \phi_j ^2)+\overline{L}_j^4 \phi_j ^4 \tan ^{-1}(\frac{x}{\sqrt{\overline{L}_j^2 \phi_j ^2-x^2}})\bigg], & \hbox{when $0\leq x<\overline{L}_j\phi_j$;} \\
                           -\frac{g_{jj}n_{j0}\phi_j^4}{16\xi_j^2}, & \hbox{when $x\geq\overline{L}_j\phi_j$.}
                         \end{array}
                       \right.\label{DoS}
\end{eqnarray}
We easily see that the strength of interspecies interaction is not contained directly in Eqs. (\ref{energyfinal}) and (\ref{DoS}), it is included in $\phi_j$ [see Eq. (\ref{profiles})].

Based on this we can calculate Casimir force, which is defined as
\begin{eqnarray}
F_C=-\frac{\partial {\mathcal E}_C}{\partial L}=-\frac{1}{2\pi}\sum_{j=1,2}\frac{\partial {\mathcal E}_{Cj}}{\partial \overline{L}_j}.\label{defineforce}
\end{eqnarray}
Combining (\ref{defineforce}) with (\ref{energyfinal}) and (\ref{DoS}) one finds
\begin{eqnarray}
F_C=\sum_{j=1,2}\frac{g_{jj}n_{j0}}{2\pi^2\xi_j^2\overline{L}_j^5}\int_0^{\overline{L}_j \phi_j } \frac{ x^3 \sqrt{\overline{L}_j^2 \phi_j ^2-x^2}}{e^{2 \pi  x}-1} \, dx.\label{force}
\end{eqnarray}
This equation shows that:

- For ideal Bose gases $g_{jj}=0$ the Casimir force is vanished as discussed in \cite{Biswas,Biswas2} for single BEC.

- In case interspecies interaction is absent (two components are miscible) $g_{12}=0$, the system behaves as the single BEC.

 We consider now the Casimir force in large inter-distance limit. In this limit, we can check that integral in (\ref{energyfinal}) get exponentially decay and is going to zero very fast. By introducing $y=x/(\overline{L}_j\phi_j)$ the first region of density of state in (\ref{DoS}) can be rewritten as
\begin{eqnarray}
\rho_j=-\frac{g_{jj}n_{j0}\phi_j^4}{8\pi\xi_j^2}\left[y(2y^2-1)\sqrt{1-y^2}+\mbox{arctan}\frac{y}{\sqrt{1-y^2}}\right],\label{large}
\end{eqnarray}
with $0\leq y<1$. Expanding (\ref{large}) in power series around $y=0$ one arrives to
\begin{eqnarray}
\rho_j\approx-\frac{g_{jj}n_{j0}\phi_j^4}{8\pi\xi_j^2}\left[\frac{8x^3}{3\overline{L}_j^3\phi_j^3}-\frac{4x^5}{5\overline{L}_j^5\phi_j^5}\right].\label{large1}
\end{eqnarray}
Plugging (\ref{large1}) into Eq. (\ref{energyfinal}) and keeping in mind that we are considering here the weakly interacting system so $\overline{L}_j\phi_j\gg1$, one is able to evaluate the integral in (\ref{energyfinal}) and result is
\begin{eqnarray}
{\cal E}_{Cj}=-\frac{m_j^3c_j^2}{360\pi\hbar^2}\left(\frac{\phi_j}{\overline{L}_j^3}-\frac{1}{7\phi_j\overline{L}_j^5}\right),\label{energyj}
\end{eqnarray}
with $c_j=\sqrt{\mu_j/m_j}$ is sound speed of condensate $j$. From Eqs. (\ref{defineforce}) and (\ref{energyj}) the Casimir force can be found in this limit
\begin{eqnarray}
F_C\approx \frac{1}{720\pi^2\hbar^2}\sum_{j=1,2}m_j^3c_j^2\left(\frac{3\phi_j}{\overline{L}_j^4}-\frac{5}{7\phi_j\overline{L}_j^6}\right).\label{Fc}
\end{eqnarray}

\begin{figure}[h]
\leavevmode
  \includegraphics[scale=0.8]{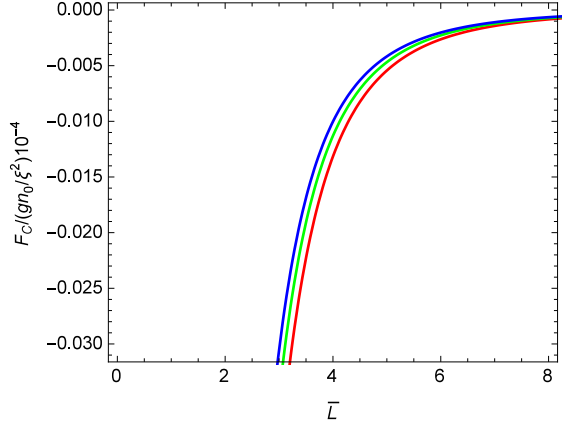}
  \caption{\footnotesize  (color online) Casimir force versus $\overline{L}$ at $K=0.5$ (red), 1 (green) and 1.5 (blue).}\label{energya}
\end{figure}

\begin{figure*}
  \mbox{
    \subfigure[\label{en2a}]{\includegraphics[scale=0.65]{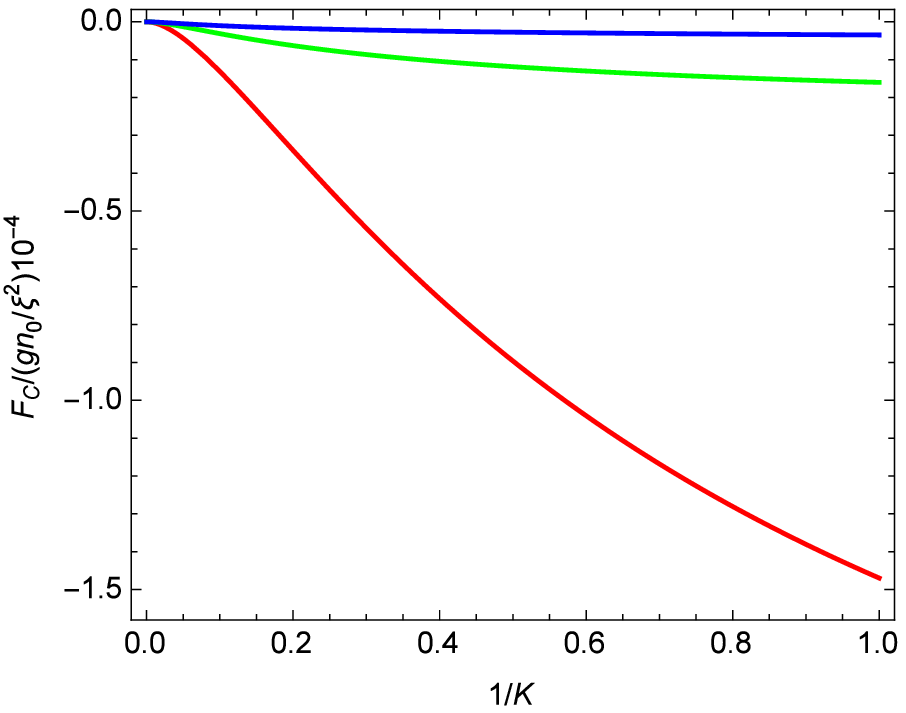}}\quad
    \subfigure[\label{en2b}]{\includegraphics[scale=0.65]{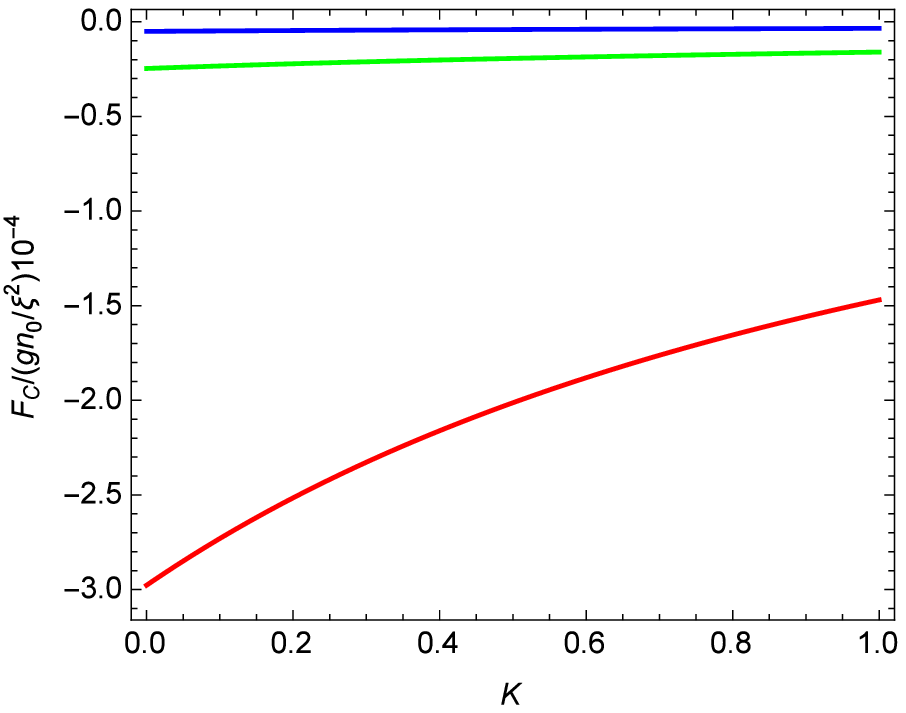}}
  }
  \caption{\footnotesize  (color online) Casimir force versus the interaction parameter at $\overline{L}=1$ (red), 2 (green), 3 (blue).}
  \label{energyb}
\end{figure*}

 In order to illustrate above calculations, we made some numerical computations. If we introduce a dimensionless quantity
\begin{eqnarray}
K=\frac{g_{12}}{\sqrt{g_{11}g_{22}}},\label{K}
\end{eqnarray}
and consider the case in which two-phase coexistence in equilibrium state, Eq. (\ref{profiles}) can be rewritten as
\begin{eqnarray}
|\phi_1|^2=|\phi_2|^2=\frac{1}{K+1}.\label{symmetry}
\end{eqnarray}
Note that for dilute gas, for single component BEC condition $na^3\ll 1$ (in their unit) has to be satisfied \cite{Andersen,Schiefele}. For binary mixtures of Bose gases we request that $n_{jj'}a_{jj'}^3\ll1$, however the definition (\ref{K}) is equivalent to
\begin{eqnarray*}
K=\frac{m_1+m_2}{2\sqrt{m_1m_2}}\frac{a_{12}}{\sqrt{a_{11}a_{22}}}.
\end{eqnarray*}
Since $m_j$ is fixed for a given system and all three scattering lengths can be changed largely independently by using the technique of Feshbach resonances \cite{Inouye}, so $K$ can run from zero to infinity for the case immiscible. For simplicity we study on the symmetric case with $g_{11}=g_{22}\equiv g, \xi_1=\xi_2\equiv\xi,n_{10}=n_{20}\equiv n_0$: in Fig. \ref{energya} one expresses the Casimir force as a function of the effective inter-distance $\overline{L}$ at several values of $K$, including miscible $K=0.5$, demixing $K=1$ and immiscible regimes $K=1.5$. We can see that as $\overline{L}$ increases the Casimir force approaches to zero very fast and there is a singular at $\overline{L}=0$; The Casimir force as a function of $1/K$ is plotted in Fig. \ref{en2a} for immiscible case and versus $K$ for miscible case in Fig. \ref{en2b}, we can see the suitability with Fig. \ref{energya}. It also shows that for full strong segregation $K\rightarrow\infty$ the Casimir force going to zero.

\section{Conclusion and outlook\label{sec:3}}

In the foregoing section we calculated the Casimir energy and Casimir force of BECs at zero temperature within framework of field theory in one-loop approximation and the periodic boundary condition is employed for $z$-direction. The our main results are in order

- Casimir energy of system is summation of the one for each component and is calculated via the density of state. For large inter-distance the Casimir energy decreases as $L^{-3}$. This result is the same as those for single component BEC.

- Casimir force is defined as the first derivative of Casimir energy with respect to plate separation. Imposing the periodic boundary condition this force is repulsive. The same as the Casimir energy, the Casimir force $F_C$ equals to summation of $F_{Cj}$ corresponding to component $j$. At large $L$ this force is decay to zero as $L^{-4}$ and this force is also vanishing in case of full strong segregation $g_{12}\rightarrow\infty$. For the case of non-interspecies interaction $g_{12}=0$ the system behaves as two distinguishing single BEC.

- Numerical computations for the Casimir force are made and show the its dependence on the parameters.

There is also an important result, Casimir force (and, of cause, Casimir energy) will be suppressed when $K$ tends to infinity and $L$ is finite. In mathematically, from Eq. (\ref{symmetry}) one can expand (\ref{force}) in power series
\begin{eqnarray}
F_c\approx \sum_{j=1,2}\frac{g_{jj}n_{j0}}{2\pi^2\xi_j^2\overline{L}_j}\left[\frac{\phi_j^4}{32}-\frac{\overline{L}_j\phi_j^5}{15}+{\cal O}(\phi_j^6)\right].
\end{eqnarray}
This equation confirms the result presented in Fig. \ref{energyb} for $1/K\rightarrow 0$. That is interesting if we can check it experimentally.

%% If you have acknowledgments, this puts in the proper section head.
\begin{acknowledgments}
%% put your acknowledgments here.
It is my pleasure to acknowledge valuable discussions with Prof. Tran Huu Phat, Jurgen Schiefele and Nguyen Thi Tham.

\end{acknowledgments}


\end{document}